# Origin of devil's staircases and anomalous Arnold tongues of breathing solitons in optical Kerr resonators

Tianhao Xian[1,2], Junsong Peng[1,2,3,*], Sonia Boscolo[4,5,6], Christophe Finot[7], Heping Zeng[1,2,†]
[1]*State Key Laboratory of Precision Spectroscopy, and Hainan Institute, East China Normal University, Shanghai 200062, China*
[2]*Chongqing Key Laboratory of Precision Optics, Chongqing Institute of East China Normal University, Chongqing 401120, China*
[3]*Collaborative Innovation Center of Extreme Optics, Shanxi University, Taiyuan, Shanxi 030006, China*
[4]*VPIphotonics GmbH, 10587 Berlin, Germany*
[5]*Aston Institute of Photonic Technologies, Aston University, Birmingham B4 7ET, United Kingdom*
[6]*Istituto Nazionale di Fisica Nucleare, Laboratori Nazionali del Sud (INFN-LNS), 95123 Catania, Italy*
[7]*Université Bourgogne Europe, CNRS, Laboratoire Interdisciplinaire Carnot de Bourgogne ICB UMR 6303, 21000 Dijon, France*
[*]*jspeng@ lps.ecnu.edu.cn* [†]*hpzeng@phy.ecnu.edu.cn*

Synchronization phenomena are ubiquitous across natural and engineered systems, and optical resonators supporting breathing solitons offer a natural platform for their investigation. In driven Kerr cavities described by the Lugiato-Lefever equation (LLE), breathing-soliton dynamics constitute a minimal setting in which such behavior can be systematically analyzed. Despite recent observations of intricate self-synchronization structures in mode-locked lasers—such as subharmonic entrainment, devil's staircases, and holes within Arnold tongues, i.e., regions of unsynchronized dynamics embedded within synchronized domains—their physical origin has remained elusive due to the complexity of laser dynamics. Here, we demonstrate that these phenomena emerge in the universal framework of the LLE. By reducing the problem to this model, we uncover the mechanism responsible for both the devil's staircase and the formation of holes within Arnold tongues. This provides a unified physical picture of synchronization in nonlinear systems, extending beyond optics and providing insight into analogous predictions and observations in fluids and other oscillatory media. In addition, we identify the regimes of high-order synchronization, enabling the generation of ultradense frequency combs, opening new directions for precision photonic applications.

The Lugiato–Lefever equation (LLE) [1] is a damped, driven, and detuned variant of the nonlinear Schrödinger equation, originally introduced to describe spatiotemporal pattern formation in dissipative, diffractive, and nonlinear optical cavities submitted to a continuous laser pump. Soon after, this equation was shown to accurately describe the dynamics observed in a dispersive optical ring cavity [2]. Since then, it has emerged as a paradigmatic model for nonlinear wave dynamics in driven-dissipative systems [3], with applications ranging from plasma physics [4] to Kerr fiber cavities [5-9] and microresonators [10-17]. From a fundamental viewpoint, the LLE has enabled the prediction and analysis of diverse intracavity structures such as solitons (bright and dark; spatial and temporal; breathers [18]), Turing patterns [19], and spatiotemporal chaos [20]. More recently, it has emerged as the central theoretical framework for broadband microresonator-based optical frequency combs, encompassing both dissipative-soliton regimes [11,21] and other nonlinear dynamical states [22-24]. Beyond classical dynamics, the quantum version of the LLE predicts nonclassical states of light such as squeezing [25], recently confirmed experimentally, opening the way to on-chip quantum photonic applications [26].

Frequency locking, or synchronization, first identified by Christiaan Huygens in 1665, is the process by which coupled nonlinear resonators adjust their frequencies to form a rational ratio, or winding number [27]. This phenomenon is ubiquitous across natural and engineered systems, arising in contexts ranging from charge-density waves [28] and Josephson junctions [29,30] to biological clocks [27], mode-locked lasers [31] and many others [32,33]. The organization of frequency-locked states in parameter space follows a number-theoretic structure described by Farey trees [31,34,35], giving rise to the devil's staircase—a self-similar fractal hierarchy of locking plateaux (periodic regions) separated by quasiperiodic regions. Such structures have been extensively studied in paradigmatic models such as the circle map [8,9] and observed across various physical systems [35-42]. A notable feature is the universality of the fractal dimension of a complete devil's staircase, approximately 0.87, independent of the underlying system [43]. The extent of synchronization is characterized by the widths of these plateaux, which grow with coupling strength and forms tongue-shaped regions—Arnold tongues [44]—in the space of frequency detuning and coupling strength. Arnold tongues belonging to different orders of synchronization can even merge with one another when the coupling strength is sufficiently strong [45,46]. These structures play a central role in controlling synchronization ranges and are therefore of direct relevance for applications [44,47,48].

Breathing-soliton solutions of the LLE provide a natural setting in which synchronization can be investigated within a minimal driven-dissipative framework. In this regime, the coexistence of the cavity repetition frequency ( $f_r$ ) and the breathing frequency ( $f_b$ ), associated with the pulse energy variation, leads to nonlinear frequency interaction and the possibility of frequency locking. Breathing solitons have been experimentally realized in Kerr fiber cavities [49], microresonators [50-54], and mode-locked lasers [31,55-57]. Periodic states (subharmonic entrainment [55,58,59]) of the breathing dynamics has been demonstrated in microresonators [59], evidencing frequency locking between $f_b$ and $f_r$ . Laser systems, where synchronization arises intrinsically without external forcing, have revealed rich synchronization behavior, including high-order Farey hierarchies and devil's staircases [31], as well as unconventional structures such as leaf- or ray-like regions and holes within Arnold tongues [57], the latter having been previously identified in theoretical studies of fluid systems and delayed-feedback microlasers [46,60]. However, the physical origin of these phenomena remains unclear due to the intrinsic complexity of laser dynamics, where dissipative processes such as gain and loss, both nonlinear, interact with conservative effects including dispersion and Kerr nonlinearity.

In this Letter, we show these synchronization phenomena within the framework of the LLE and develop an analytical description to capture their origin. Our approach is based on a discrete-time formulation of the LLE, in which the continuous evolution over one roundtrip is recast as an iterative map. Within this framework, we derive reduced equations governing the evolution of breathing-soliton parameters, establishing a direct connection between the LLE and the circle map. This provides a physical interpretation of the devil's staircase and reveals the holes within Arnold tongues originating from a sinusoidal dependence of the breathing frequency on the control parameter. This mechanism further enables access to extremely high-order synchronization states within Arnold holes, with a regime identified at winding number of 99/439. From an applications perspective, the spacing of frequency combs in Kerr cavities is fundamentally determined by the cavity free spectral range, leading in microresonator implementations to relatively sparse comb spectra [38,52]. Our work demonstrates that this limitation can be overcome by exploiting breathing-soliton dynamics, thereby enabling significantly denser comb generation. The analytical predictions show good agreement with full numerical simulations.

Although breathing-soliton solutions of the LLE have been widely investigated, their synchronization dynamics remain poorly understood. The mean-field LLE is averaged over roundtrips and does not retain the full discrete boundary condition, although its slow time remains connected to the cavity roundtrip number and the breathing dynamics can be sampled stroboscopically once per roundtrip. Here, we overcome this limitation by adopting a discrete-time formulation that explicitly retains the roundtrip-to-roundtrip evolution, allowing the synchronization between the breathing frequency and the cavity repetition frequency to be described through an iterative map. We consider the LLE in the form [61]:

$$\frac{\partial E}{\partial t} = \kappa\left( i|E|^2 E + \frac{i}{2}\frac{\partial^2 E}{\partial \tau^2} - E + E_{\mathrm{in}} e^{i\kappa\alpha t} \right), \tag{1}$$

where $t$ is the slow time that describes the evolution of the field envelope $E(t,\tau)$ at the scale of the cavity roundtrip time, $\tau$ is a fast time defined in a reference frame that travels at the group velocity of light, $E_{\mathrm{in}}$ the driving amplitude, $\kappa$ the cavity loss, and $\kappa\alpha$ the detuning. This equation can be transformed into the standard normalized LLE form [1] through appropriate rescaling. Writing the driving term in exponential form provides a compact representation of the coherent pump in the rotating frame. This structure naturally separates nonlinear propagation and linear (loss and driving) contributions, leading to an iterative map in which each step corresponds to one cavity roundtrip [see the inset of Fig. 1(a) for the system schematic and Supplementary Material (SM) for the full derivation].

An approximate solution of Eq. (1) is constructed as a superposition of a continuous-wave (cw) background and a soliton contribution [4,10], $E(t,\tau) \simeq E_{\mathrm{cw}}(t) + NE_s(t,\tau)e^{i\kappa\alpha t}$, $E_s(t,\tau) = \eta\,\mathrm{sech}(\eta\tau)\, e^{-i\chi - i\pi/2}$, where $E_{\mathrm{cw}}$ is the homogeneous background, $E_s$ the soliton profile, $\eta$ characterizes its amplitude and duration, and $\chi + \pi/2$ is the phase relative to the cw background. The soliton component is represented by the fundamental soliton profile $E_s$, with $N$ denoting its soliton-order parameter [62]. Within the discrete-time framework, soliton perturbation theory [4,61,63] yields recursion relations for the soliton parameters,

$$\begin{aligned} \eta_{n+1} &= \eta_n + \kappa\eta_n\left[2\cos\left(\alpha - \delta\eta_n^2\right) - 4\right] - \frac{\pi}{2} l E_{\mathrm{in}} \kappa \eta_n^2 \cos\chi_n \\ \chi_{n+1} &= \chi_n + \kappa\left[\sin\left(\alpha - \delta\eta_n^2\right) - \frac{\eta_n^2}{2}\right] + \pi l E_{\mathrm{in}} \kappa\eta_n \sin\chi_n \end{aligned}, \tag{2}$$

where $n$ denotes the roundtrip index, $l(\kappa,\alpha) = |E_{\mathrm{cw}}|/E_{\mathrm{in}}$ , and $\delta = N^2 - 1$ .

To describe the breathing dynamics, we represent the soliton parameter in slow time as a circular modulation about its stationary value, $\eta(t)e^{-i\chi(t) - i\pi/2} \simeq \eta_0 e^{-i\chi_0 - i\pi/2} + re^{-i\vartheta(t)}$, where $\eta_0 e^{-i\chi_0 - i\pi/2}$ denotes the stationary soliton parameter and $re^{-i\vartheta(t)}$ its modulation. Linearizing the soliton-parameter recursion relations about the stationary state and transforming to the corresponding phase variable yields the following discrete-time phase map (see the SM for the derivation):

$$\varphi_{n+1} = \varphi_n + \omega + \rho \sin \varphi_n . \tag{3}$$

The resulting map is the canonical circle map, which provides a standard framework for synchronization and the emergence of devil's staircases [34,43,64-66]. Here, $\omega$ is the effective drift of the breathing phase and $\rho$ is the coupling strength. The phase-dependent term, whose amplitude is quantified by $\rho$, defines the coupling between the breathing frequency and the repetition frequency, which arises from the phase-dependent roundtrip evolution of the soliton parameters. This coupling reflects the fact that the phase of the breathing oscillation relative to the cavity driving affects the perturbation accumulated during each roundtrip, due to the non-homogeneous distribution of pump injection and attenuation within the roundtrip. The effective drift and coupling strength are given by

$$\omega = 2\kappa\left(\sin q_0 - \eta_0^2 + \tfrac{5}{4}P\eta_0 \sin\chi_0 - \delta\eta_0^2\cos q_0\right), \tag{4}$$

$$\rho = 2\sqrt{A_\vartheta^2 + B_\vartheta^2}, \tag{5}$$

where $q_0 = \alpha - \delta\eta_0^2$, $P = \pi l E_{\text{in}}$, and $A_\vartheta$ and $B_\vartheta$ are the coefficients arising from the linearized breathing dynamics [see Supplementary Eq. (S12)]. In the absence of the phase-dependent coupling term, Eq. (4) gives the free-running breathing frequency, $\omega/2\pi$. Equation (3) therefore captures its nonlinear interaction with the cavity repetition frequency and the resulting synchronization dynamics. Equations (3)–(5) show that the locking range can be controlled through the cavity loss, pump strength, and detuning, which is important for the realization of frequency locking.

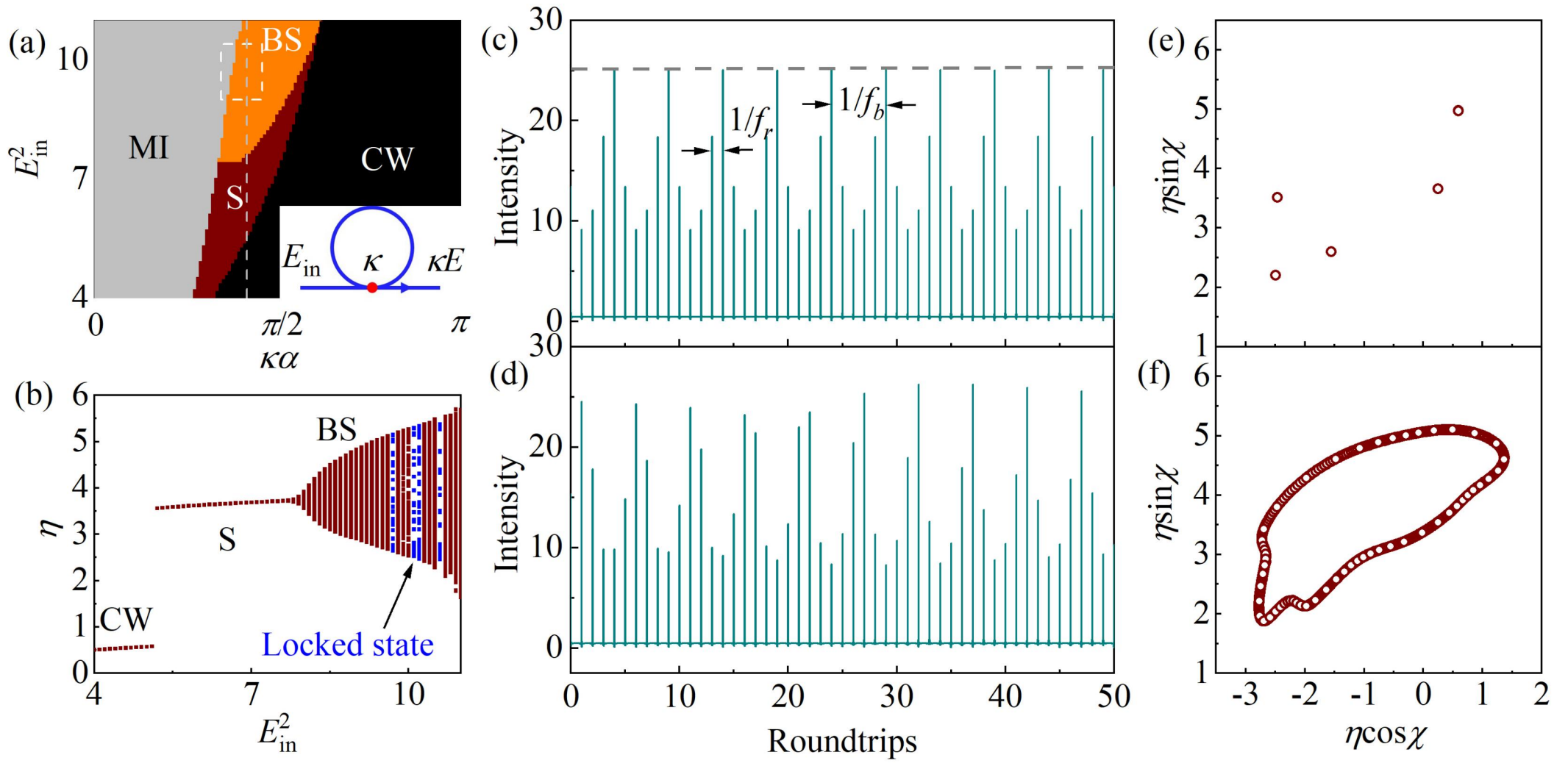

FIG. 1. Synchronized and desynchronized Kerr breathing solitons. (a) Phase diagrams of LLE in the ($\alpha$, $E_{\text{in}}^2$) plane at $\kappa = \pi/12$. The inset shows a schematic of the ring-cavity system studied in this work. The white rectangular marks the location of Fig. 3(a). (b) Hopf bifurcation for $\alpha = 5$ [gray dashed line in (a)]. (c, d) Intensity profiles of synchronized and unsynchronized breathing solitons, respectively, over consecutive cavity roundtrips; the pulse repetition period $1/f_r$ and the breathing period $1/f_b$ are marked in (c). (e, f) Corresponding Poincaré sections in the ($\eta$, $\chi$) plane over 5000 roundtrips, showing a discrete period-5 orbit in the synchronized regime and a closed curve in the desynchronized regime.

Numerical simulations are performed using an iterative scheme derived from Eq. (1), in which roundtrip propagation and boundary conditions are treated separately. By varying the detuning $\alpha$ and the driving strength $E_{\text{in}}$, the system exhibits distinct dynamical regimes, including stationary solitons (S), breathing solitons (BS), continuum wave (CW), and modulation instability (MI), as illustrated in Fig. 1(a). Consistent with previous studies [52-54], breathing solitons arise in the intermediate regime between stationary solitons and MI, with their onset corresponding to a Hopf bifurcation [Fig. 1(b)]—a universal mechanism observed across a wide range of nonlinear systems, including optomechanical, fluid, and optical resonators [18,51,52,67-72]. Breathing solitons may display either synchronized (periodic) or desynchronized (quasiperiodic) dynamics with respect to the cavity roundtrip, as reported in prior works [31,55,59]. In the synchronized regime, the soliton exhibits strictly periodic energy oscillations, as illustrated in Fig. 1(c), whereas in the desynchronized regime this periodicity is degraded [Fig. 1(d)]. This distinction is further elucidated in the Poincaré section spanned by the soliton parameters ($\eta$, $\chi$), shown in Fig. 1(e, f), where the synchronized and desynchronized states correspond to a discrete periodic orbit and a closed curve, respectively. In particular, Fig. 1(e) displays five distinct points, consistent with a periodicity of five roundtrips in Fig. 1(c), corresponding to a winding number $f_b/f_r = 1/5$.

Figure 2(a) shows the scaling of the winding number with the cavity loss $\kappa$, revealing an overall linear dependence interrupted by a sequence of plateaus (red). These plateaux correspond to synchronized states, where the winding number remains constant over finite intervals of $\kappa$, while the intervening linear segments correspond to desynchronized dynamics. The global linear trend (green) is well captured by Eq. (4), which describes the scaling of the free-running breathing frequency with $\kappa$. In the fitting, $l$ in Eq. (4) is set to its value at $\kappa$=0.213 ($l = 0.22$), $\eta_0 = 3.50$ and $\chi_0 = 1.81$ are obtained from the mean value of the complex soliton parameter, $\langle \eta e^{-i\chi - i\pi/2} \rangle$, over 5000 roundtrips, and $\delta$ is set to zero. The fit agrees well with the simulation results, with a slope of 1.04 and a small intercept error of –0.026, indicating good agreement between the reduced model and the LLE simulations. The plateaux emerge sequentially following the Farey tree ordering [31,35]: given two adjacent elements $m$:$n$ and $p$:$q$ of the Farey sequence (parent plateaux), a new plateau appears at their Farey sum ($m$+$p$):($n$+$q$). This hierarchy is illustrated in the inset of Fig. 2(a), where, for example, the relation 1:5 ⊕ 1:4=2:9 is satisfied. Importantly, regions that initially appear devoid of plateaux—such as those highlighted by dashed boxes—reveal additional synchronization intervals when the resolution in $\kappa$ is increased, as shown in Figs. 2(b) and 2(c). This self-similar proliferation of plateaux indicates that their number is unbounded, a defining feature of the fractal structure known as the devil's staircase.

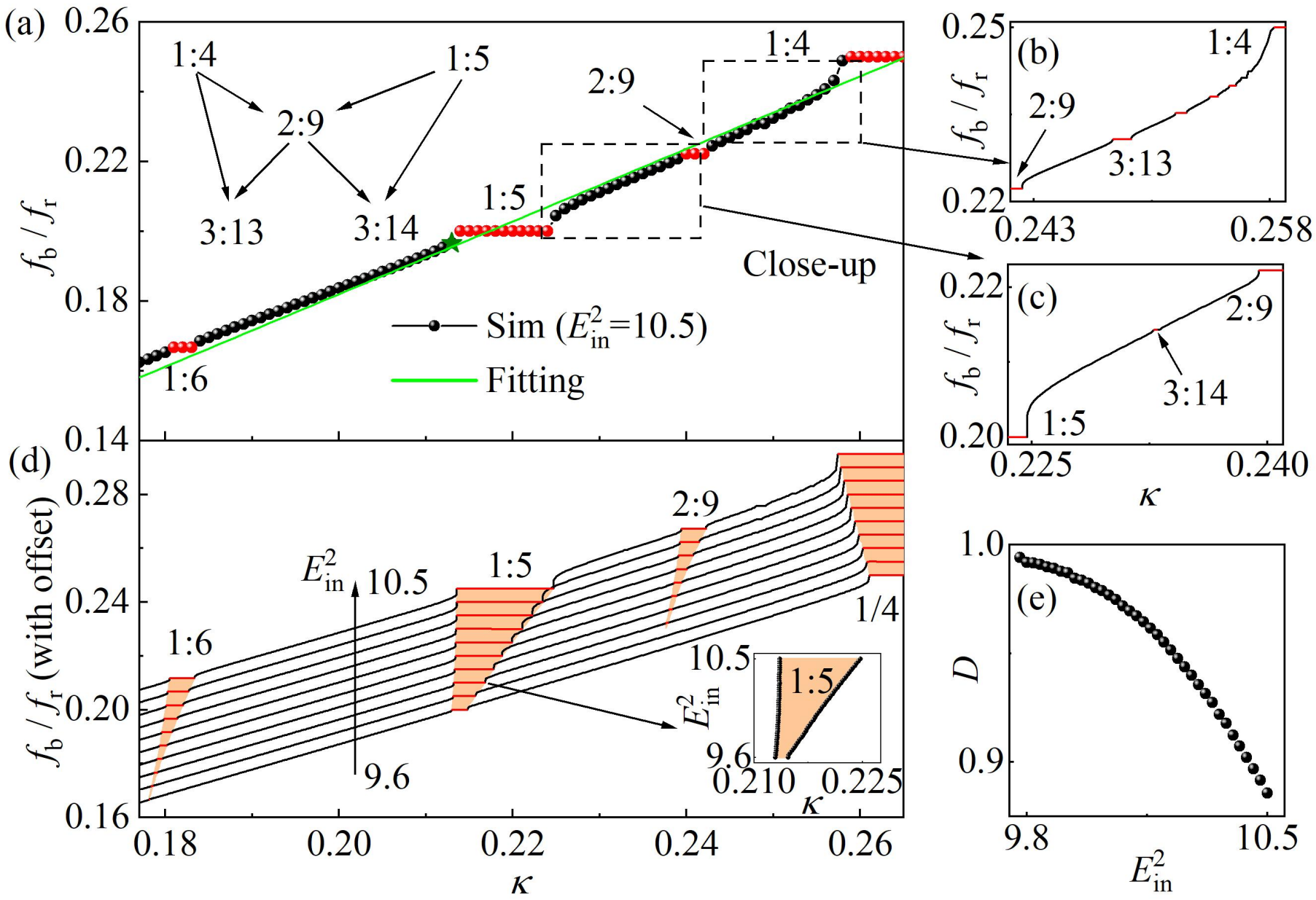

FIG. 2. Devil's staircase and Arnold tongue structure. (a) Winding number $f_b/f_r$ as a function of cavity loss $\kappa$ (step size 0.001), showing a devil's staircase (red) superimposed on a linear trend predicted by Eq. (4). The green line represents the linear fit based on Eq. (4), yielding a slope of 1.04 and an intercept of -0.026. The inset illustrates the corresponding Farey tree hierarchy. The star marks the point used to calculate the parameters for the fit. (b, c) Enlarged views of the regions highlighted in (a), obtained with finer resolution in $\kappa$ (step size 0.0001), revealing additional plateaus and confirming the fractal structure. (d) Waterfall plot of the winding number versus $\kappa$ for increasing driving strength $E_{in}^2$ (from 9.6 to 10.5), with a vertical offset of 0.005 for clarity. Synchronized regions are shaded, forming tongue-like structures; an example for winding number 1/5 is shown in the inset. (e) Fractal dimension $D$ of the staircase as a function of driving strength $E_{in}^2$.

A central question is whether these synchronization regions follow the Arnold tongue structure. In externally driven systems, Arnold tongues are readily mapped since the coupling strength between resonators can be directly controlled. In contrast, here the interaction occurs between two intrinsic frequencies, the breathing frequency $f_b$ and the cavity repetition frequency $f_r$, making the effective coupling less direct to access. Nevertheless, Eq. (5) shows that the coupling parameter $\rho$ can be tuned via the driving strength $E_{in}$. This is confirmed in Fig. 2(d), where the width of the synchronization plateaus increases with $E_{in}^2$. The vertical offset is introduced for clarity. To explicitly demonstrate the Arnold tongue structure, the synchronization region corresponding to a winding number of 1:5 is mapped in the parameter space $(\kappa, E_{\rm in}^2)$, as shown in the inset of Fig. 2(d), exhibiting the characteristic tongue-like shape. Additional mappings for the synchronization regions associated with winding numbers 1:6 and 2:9 are provided in Fig. S1 of SM.

The fractal dimension $D$ of the staircase is evaluated using the scaling relation $S_1^D + S_2^D = S^D$ [35,73], where $S_{1,2}$ denote the gaps between a child plateau and its parent plateaux, and $S$ is the gap between the two parent plateaux. In the weak-coupling regime (small $E_{\rm in}^2$), the coupling amplitude $\rho$ [Eq. (5)] remains small, and the scaling relation yields $D \to 1$. As the coupling strength increases, $D$ decreases, reaching ≈ 0.886 at $E_{\rm in}^2$ = 10.5, in close agreement with the universal value 0.87 associated with a complete devil's staircase [8]. This evolution is summarized in Fig. 2(e).

Together, these results provide clear evidence of a devil's staircase structure in the LLE system. This behavior originates from the reduction of the breathing-soliton dynamics to the circle map [Eq. (3)], which is known to exhibit such fractal synchronization patterns [34,43,64,74]. Similar staircase structures are observed at other fixed $\alpha$ (see Supplementary Fig. S2), supporting the robustness of the phenomenon. Similar structures are also observed when varying the driving strength (see Fig. S3 in the SM).

Distinct synchronization dynamics emerge when varying the detuning parameter $\alpha$ at different driving strengths $E_{\rm in}^2$, as mapped in the ($\alpha$, $E_{\rm in}^2$) parameter space in Fig. 3(a) and illustrated by several curves in Fig. 3(b). Three key features differentiate this behavior from the $\kappa$-scaling in Fig. 2(a): (i) the winding-number curves exhibit pronounced bending rather than a global linear trend; (ii) synchronization plateaux are interrupted by desynchronized intervals [arrows in Fig. 3(b)], corresponding to holes within Arnold tongues [57,60]; and (iii) the synchronization regions deviate from the conventional tongue-like shape, as exemplified by the plateaux with winding numbers 2:9, 9:40, 7:31, and 5:22, which broaden and subsequently narrow as pump intensity increases.

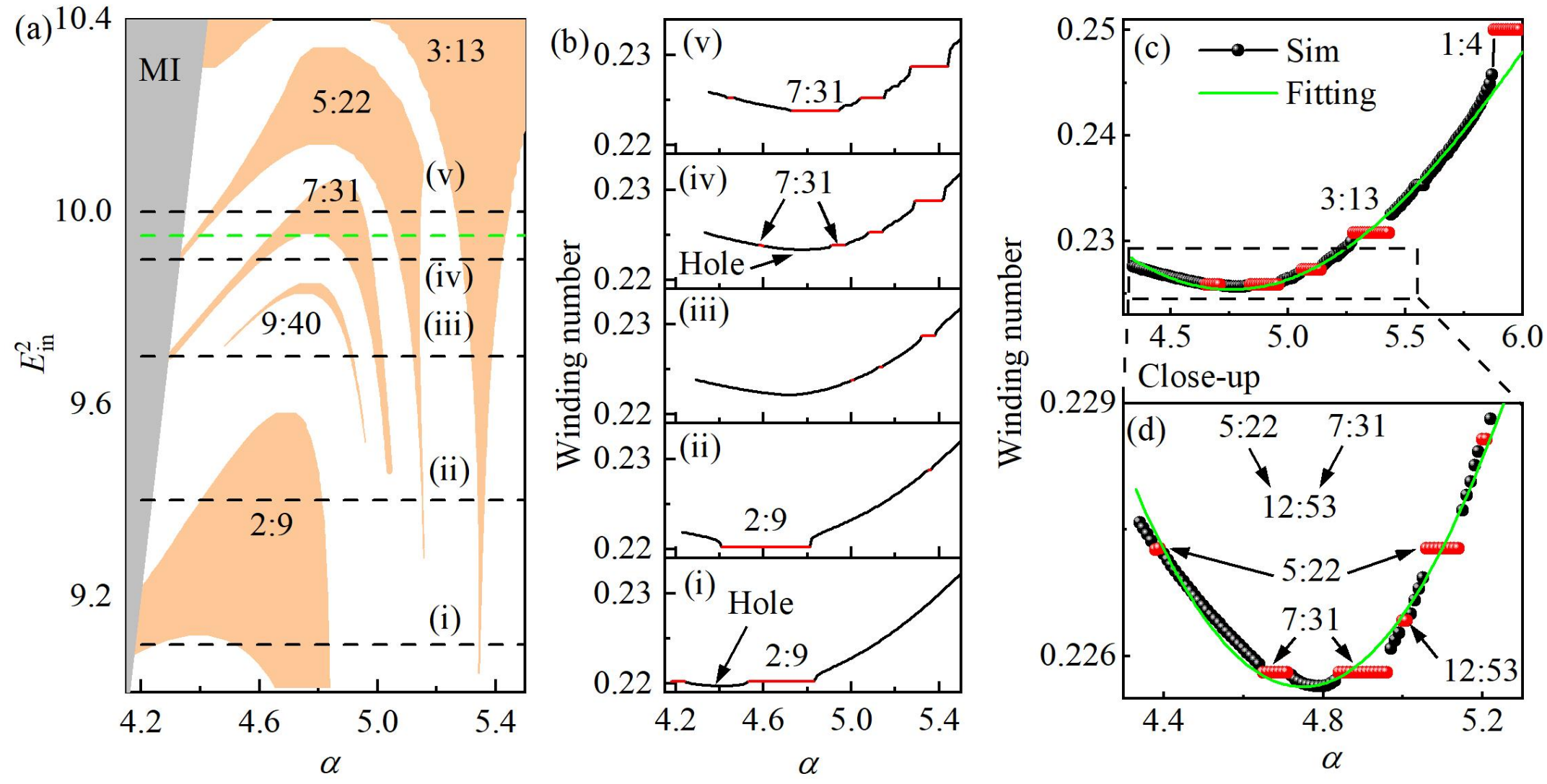

FIG. 3. Anomalous synchronization dynamics: Arnold tongue holes and non-tongue shape of the synchronization regions. (a) Synchronization map at cavity loss $\kappa = \pi/12$. Locking regions associated with rational winding numbers are shown in brown, while the shaded gray region indicates modulation instability. (b) Winding number $f_b/f_r$ as a function of detuning $\alpha$ (step size 0.01), marked as black dashed line in (a). The red lines highlight the plateaux. (c) Winding number versus $\alpha$ at $E_{\rm in}^2$ = 9.95 [green dashed line in (a)], showing agreement with the sinusoidal dependence predicted by Eq. (4). The green curve represents a sinusoidal fit to the numerical data with arbitrary fitting coefficients. (d) Magnified view of the region indicated in (c); the inset shows the corresponding Farey construction. The plateaux in (c) and (d) are colored red.

These features are consistently explained by Eq. (4). As illustrated in Fig. 3(c), which shows a representative curve [green trace in Fig. 3(a)], the winding-number dependence closely follows the sinusoidal dependence predicted by Eq. (4). The symmetry of this dependence plays a central role: the formation of a plateau on one side of the curve is accompanied by a corresponding structure on the opposite side, leading to the appearance of desynchronized gaps within otherwise continuous synchronization regions, i.e., holes in Arnold tongues. This mechanism is further clarified in Fig. 3(d), where a magnified region reveals that the plateau at 7:31 is interrupted by desynchronized states;

similar behavior is observed for the 5:22 plateau. Between them, a child plateau at 12:53 emerges, forming a Farey branch as shown in the inset. The asymmetry of the plateaux is also evident: synchronization regions are broader for $\alpha$>4.8 than for $\alpha$<4.8, reflecting a stronger effective coupling in this regime. Consequently, the 12:53 plateau is clearly visible on the right side, while it is suppressed on the left at the present resolution; it reappears when a finer sampling in $\alpha$ is used (see Fig. S4 in SM). Although holes in Arnold tongues have been reported previously [28,29], their physical origin has remained unclear; here, they are directly traced to the sinusoidal dependence encoded in Eq. (4).

We now turn to the non-tongue shape of the synchronization regions. According to Eq. (4), an enlargement of $E_{\text{in}}^2$ makes the free-running breathing frequency ω increase, thus increasing the winding number. The edges of a hole, at which the winding number is smaller than that of the plateau, can become locked to the plateau, as illustrated for the 2:9 and 5:22 plateaux in Fig. 3(a). This process leads to broadening of the synchronization region, which reaches a maximum when the hole is completely closed. Upon further increase of $E_{\text{in}}^2$, the winding number at the plateau boundaries increases and deviates from the synchronized state, resulting in progressive narrowing and eventual disappearance of the synchronization region. Related mechanisms have also been reported for the merging, connecting, and disconnecting of resonance tongues [46,75].

In addition, Figs. 3(a) and 3(b) shows that the detuning value at which $f_{\text{b}}/f_{\text{r}}$ attains its minimum shifts to larger $\alpha$ as $E_{\text{in}}^2$ increases. This behaviour is also captured by Eq. (4), which shows that the winding number has an oscillatory dependence on the detuning through $q_0$. The stationary quantities $\eta_0$ and $\chi_0$also vary with $\alpha$, so the extrema of the full winding-number curve are determined by their combined dependence rather than by the explicit sinusoidal terms alone. Nevertheless, the underlying sinusoidal dependence gives rise to the characteristic oscillatory structure of the winding-number curve, while the positions of its extrema are modified by the $\alpha$-dependence of the stationary pulse parameters. As the driving strength increases, the pulse energy grows, leading to larger $N$ and hence a systematic change in $\delta = N^2 - 1$ and in the stationary pulse parameters, which shifts the positions of the extrema.

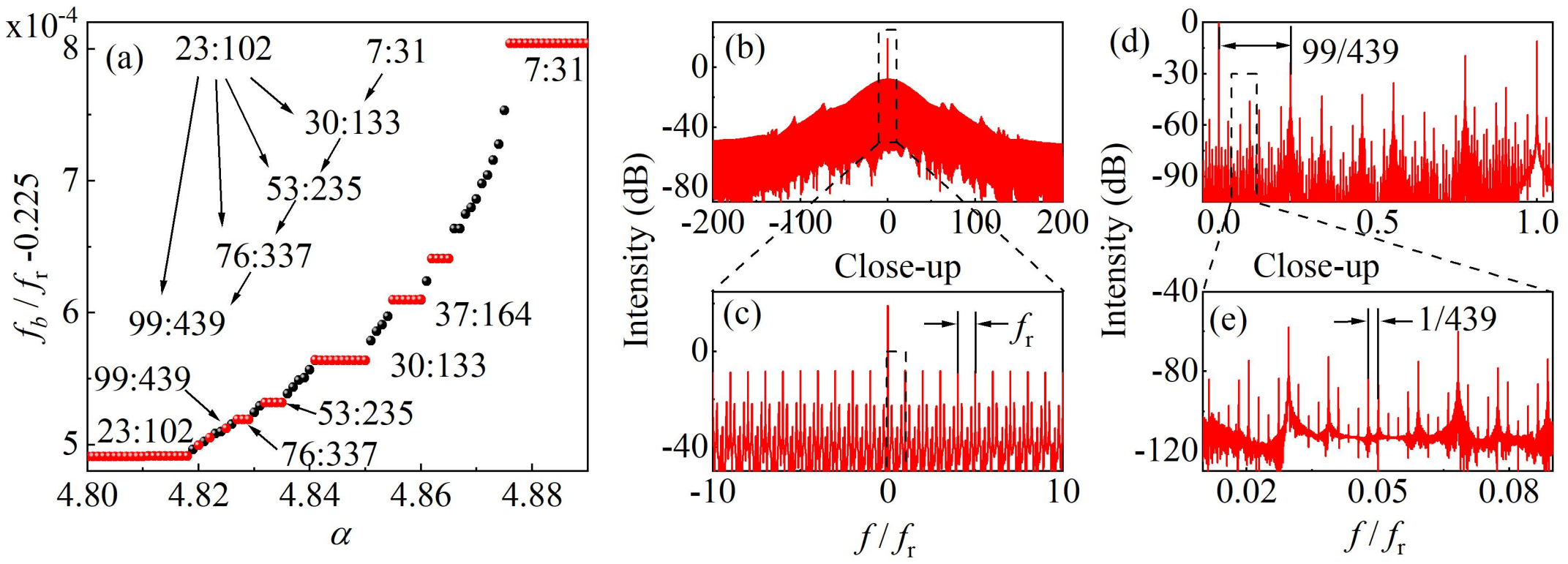

FIG. 4. High-order synchronization near the extremum of $f_{\text{b}}/f_{\text{r}}$. (a) Winding number $f_{\text{b}}/f_{\text{r}}$ as a function of detuning $\alpha$, showing a dense hierarchy of synchronization plateaux near the extremum; the inset displays the corresponding Farey tree structure. At each point, the simulation is performed over 1.1x10$^5$ roundtrips, with the first 10$^4$ roundtrips discarded for the calculation of the winding number. The locked states are further verified using Poincaré sections (see Supplementary Fig. S5). (b) Optical frequency spectrum of the highest-order synchronized state with winding number 99:439, calculated over $2^{17}$ roundtrips. By taking the roundtrip time $T_r$ as the simulation time window, we construct a time series containing $2^{29}$ points using $T = \tau + nT_r$, where $n$ denotes the roundtrip index and each roundtrip contains $2^{12}$ temporal points, from which the spectrum is obtained. (c) Magnified view of the region highlighted in (b), resolving the optical comb lines separated by one free spectral range (FSR, $f_{\text{r}}$). (d) Optical spectrum within one FSR, revealing the dense spectral structure generated by the high-order synchronized state. The location of (d) in (c) is marked by the black dashed rectangle. (e) Magnified view of the region highlighted in (d), resolving the individual spectral lines with spacing 1/439.

The presence of an extremum in the winding number $f_{\text{b}}/f_{\text{r}}$ implies that synchronization is favored in its vicinity, where the variation of $\omega$with respect to the control parameter $\alpha$is reduced. At the extremum, the first derivative of $\omega(\alpha)$ vanishes, giving a locally quadratic dependence on $\alpha$. Consequently, the locking interval in the control parameter is broadened from $O(\rho)$ to $O(\sqrt{\rho})$. This enhanced robustness of phase locking allows synchronization to persist for closely spaced rational frequencies, thereby enabling high-order states with large denominators. As shown in Fig. 4(a), a dense hierarchy of synchronization states is observed, with the largest denominator reaching 439— nearly an order of magnitude higher than values reported in previous physical systems [39-42,76,77]. Poincaré sections in the $(\eta, \chi)$ plane for this highest-order synchronized state and a nearby desynchronized state are presented in Fig. S5 of SM. Arnold's tongues in the $(\alpha, E_{\mathrm{in}}^2)$ plane of several plateaux in Fig. 4(a) are presented in Fig. S6 based on experimentally realistic parameter steps, verifying the experimental observability of these periodic states. The convergence of the devil's staircase with respect to simulation time, numerical resolution, and initial conditions is further demonstrated in Fig. S7 of SM. While finer steps likely exist due to the fractal nature of the devil's staircase, this "record" value reflects the resolution of our simulations. Experimentally, accessing such high-order steps would require fine control of $\alpha$ and $E_{\mathrm{in}}^2$, and the corresponding states may be narrower and less robust than lower-order steps.

Furthermore, the simulations illustrate that high-order synchronization can dramatically increase comb line density. Figures 4(b–e) illustrate the resulting optical spectral structure. For the synchronized state with winding number 99:439, 438 breathing-induced spectral lines appear within each free spectral range, in addition to the soliton comb line, with a spacing of $f_{\mathrm{r}}/439$. Thus, the optical spectral-line density is 439 times higher than that of the stationary soliton over the same frequency interval. The generated lines are equally spaced and commensurate with the underlying soliton comb structure. In the noise-free simulations, these lines are discrete, reflecting the periodicity of the synchronized state. Experimental linewidth and mutual coherence, however, will depend on intrinsic and technical noise and require separate characterization.

These results demonstrate that high-order synchronization provides an efficient mechanism to enhance the line density of microresonator-based frequency combs, a capability particularly relevant for applications requiring densely spaced spectral channels, including high-capacity optical communications and massively parallel ranging.

In conclusion, we have demonstrated that breathing solitons in Kerr cavities governed by the LLE exhibit a devil's staircase structure and unconventional synchronization dynamics, including the formation of holes within Arnold tongues and non-monotonic synchronization regions. By formulating the LLE as a discrete roundtrip map and applying soliton perturbation theory, we have established a direct connection between breathing-soliton dynamics and the circle map, providing a unified physical interpretation of these phenomena. These results offer insight into analogous synchronization behaviours observed in the more complex mode-locked lasers [31,55-57] and in fluid flows [29], while highlighting the LLE as a minimal and tractable platform for their investigation. Beyond the single-soliton regime considered here, Kerr cavities can support multiple breathing solitons [52], where each pulse behaves as an individual resonator. This suggests that multi-soliton states may provide a natural setting to explore collective synchronization phenomena within a single resonator, analogous to those reported in coupled resonator networks [78]. Furthermore, the present framework can be extended to systems of coupled Kerr resonators [79,80], where the dynamics may be described in terms of coupled circle maps [81], offering a route to understanding complex synchronization patterns in interacting nonlinear systems. Finally, since devil's staircase structures are known to transition to chaos as the effective coupling increases, the LLE provides a platform to explore a distinct route to broadband chaos generation. In contrast to modulation-instability-driven chaos [82], chaos originating from breathing solitons is inherently broadband, potentially enabling a larger number of parallel channels for applications such as optical ranging [83,84].

**Acknowledgments:** We acknowledge support from the Innovation Program for Quantum Science and Technology (2023ZD0301000), the National Natural Science Fund of China (62305209, 12434018).
**Declaration of competing interests:** The authors declare no competing interest.
**Data availability:** Data will be made available on request.